1

# Ion heating and magnetic flux pile-up in a magnetic reconnection experiment with super-Alfvénic plasma inflows

L. G. Suttle,[1,a)] J. D. Hare,[1] S. V. Lebedev,[1,b)] A. Ciardi,[2] N. F. Loureiro,[3] G. C. Burdiak,[1,c)] J. P. Chittenden,[1] T. Clayson,[1] J. W. D. Halliday,[1] N. Niasse,[1,c)] D. Russell,[1] F. Suzuki-Vidal,[1] E. Tubman,[1] T. Lane,[4] J. Ma,[5] T. Robinson,[1] R. A. Smith,[1] N. Stuart[1]

[1]*Blackett Laboratory, Imperial College, London, SW7 2BW, United Kingdom*
[2]*Sorbonne Université, Observatoire de Paris, Université PSL, CNRS, LERMA, F-75005, Paris, France*
[3]*Plasma Science and Fusion Center, Massachusetts Institute of Technology, Cambridge, Massachusetts, 02139, USA*
[4]*West Virginia University, Morgantown, West Virginia 26506, USA*
[5]*Northwest Institute of Nuclear Technology, Xi'an 710024, China*

(Revised January 26, 2018)

This work presents a magnetic reconnection experiment in which the kinetic, magnetic and thermal properties of the plasma each play an important role in the overall energy balance and structure of the generated reconnection layer. Magnetic reconnection occurs during the interaction of continuous and steady flows of super-Alfvénic, magnetized, aluminum plasma, which collide in a geometry with two-dimensional symmetry, producing a stable and long-lasting reconnection layer. Optical Thomson scattering measurements show that when the layer forms, ions inside the layer are more strongly heated than electrons, reaching temperatures of $T_i \sim \bar{Z}T_e \gtrsim 300$ eV – much greater than can be expected from strong shock and viscous heating alone. Later in time, as the plasma density in the layer increases, the electron and ion temperatures are found to equilibrate, and a constant plasma temperature is achieved through a balance of the heating mechanisms and radiative losses of the plasma. Measurements from Faraday rotation polarimetry also indicate the presence of significant magnetic field pile-up occurring at the boundary of the reconnection region, which is consistent with the super-Alfvénic velocity of the inflows.

## I. INTRODUCTION

The interaction of magnetized plasma flows occurs in many astrophysical systems (e.g. stellar jets[1], supernovae[2], accretion disks[3]), space environments (e.g. solar flares, solar wind-magnetosphere interactions)[4] and high energy density laboratory experiments (e.g. laser-plasma interactions[5,6], Z-pinches[7] and inertial confinement fusion[8,9]). For colliding plasma flows with oppositely-directed, embedded magnetic fields, the reversal of the field direction across their interface gives rise to a current sheet. In this layer, the frozen-in flux condition breaks down and the plasma and magnetic field decouple, allowing the field lines to break and reconnect, releasing stored magnetic energy. The spatial scale of the reconnection layer is controlled by the interplay of the plasma resistivity, two-fluid and kinetic effects, which drive this transition from ideal magnetohydrodynamic behavior[10–12]. This process depends strongly on the external boundary conditions, and the structure of the reconnection layer adjusts to balance the magnetic and material fluxes brought to the reconnection region, and the rate of magnetic annihilation and the outflow of energized material[13]. This is apparent if the plasma flow into the reconnection region is strongly driven, such that the ram pressure of the material flux is significant in comparison to the magnetic pressure. A number of recent laser-driven, high energy density physics (HEDP) experiments have investigated magnetic reconnection in conditions where the ram pressure is much higher than the magnetic pressure[14–19]. In those experiments, the interaction of expanding plasma plumes from solid targets irradiated by 1-2 ns duration laser pulses results in the transient annihilation of thin sheets of toroidal magnetic fields.

This paper presents data from experiments carried out on a recently developed pulsed power reconnection platform[20–23], which applies a 1 MA, ~500 ns current pulse to an array of thin wires to produce high velocity, counter-streaming plasma flows with oppositely-directed, embedded magnetic fields. An important feature of this platform is that the reconnection layer is long-lasting, as it is continuously supported by the inflowing magnetized plasma for the duration of the experiment. This allows sufficient time for the density and magnetic field structures to form and evolve. The geometry of the layer displays a two-dimensional symmetry, which allows for good diagnostic access. The setup also offers a versatile testbed for studying magnetic reconnection over a broad range of plasma conditions, as it

---

a) Email: l.suttle10@imperial.ac.uk
b) Email: s.lebedev@imperial.ac.uk
c) Current address: First Light Fusion Ltd, Oxfordshire, OX5 1QU, United Kingdom

is possible to control the inflow properties via the choice of the plasma material. For example, the use of either aluminum or carbon wires produces flows with a super[20] or sub-Alfvénic velocities[21–23] respectively. The reconnection layers formed in experiments with these two materials have different Lundquist numbers[13] ($S=LV_A/D_M$, where L is the length scale of the plasma, $V_A$ is the Alfvén speed of the upstream plasma and $D_M$ is the magnetic diffusivity), allowing access to different regimes of magnetic reconnection parameter space (e.g. single vs. multiple x-line reconnection)[24]. The work presented in this paper extends results previously published in Ref. 20, and describes the detailed characterization of a reconnection layer formed in an aluminum plasma, with strongly driven inflows ($M_A=V_{flow}/V_A\approx 2$). The Lundquist number for the layer is relatively small ($S\sim 10$) due to strong radiative cooling of the aluminum plasma, which limits the electron temperature. Characterization of the reconnection layer structure is made via detailed spatially and temporally resolved, quantitative measurements of the plasma parameters, obtained using a comprehensive suite of diagnostics.

The paper is organized as follows. Sec. II describes the setup of the pulsed power magnetic reconnection platform, which uses a wire array configuration to produce sustained, counter-streaming, magnetized plasma flows. It also describes the diagnostic setup of high-speed optical imaging, laser interferometry, optical Thomson scattering and Faraday rotation polarimetry, which are used to make non-perturbative measurements of the plasma. Sec. III presents the results, showing the conditions of the plasma following the formation of the reconnection layer, and describes how the plasma parameters evolve over time. These measured parameters are summarized in Table I in Sec. IV, and this is accompanied by a discussion of the main features of the reconnection layer structure. In this section a brief comparison is also made to reconnection occurring in experiments with sub-Alfvénic carbon plasma flows[21,22] at a much larger Lundquist number of $S\sim 100$ (for a more in depth comparison see the review article of Ref. 23). The conclusions of this work are summarized in Sec. V.

## II. EXPERIMENTAL SETUP AND DIAGNOSTICS

The experiments were carried out at the MAGPIE pulsed power facility[25], using the setup illustrated in Fig. 1(a). The supersonic, counter-streaming plasma flows are produced by the ablation of thin aluminum (Al) wires, driven by a 1 MA, ~500 ns current pulse. These wires are arranged to form two cylindrical, "inverse" wire arrays[26], with the total current divided equally between the two arrays. The current in each array runs up the wires and down the central conductor, as indicated by the (purple) arrows in Fig. 1(a). Plasma is continuously ablated from the resistively heated wires, and the J×B force acts on the plasma driving supersonic plasma flows, which are sustained throughout an entire experiment. This is similar to the ablation plasma flows produced by standard (Z-pinch) wire arrays[27–29], however, here the J×B force acts to direct the plasma radially outwards, into a region initially free of magnetic fields. The ablated plasma is accelerated away from the wires within the first 1-2 mm, and thereafter propagates with an almost constant velocity[26]. Previous measurements have demonstrated that the plasma flows generated by a single, inverse Al wire array have a frozen-in, azimuthal, advected magnetic field (B~2 T), and the velocity of the flows is super-fast-magnetosonic (i.e. $V_{flow} > V_{FMS} = [c_S^2 + V_A^2]^{1/2}$, where $c_S$ is the ion sound speed)[30–32]. The arrays used in the current experiments consist of 16 Al wires, each 30 μm in diameter and 16 mm in length, and positioned with uniform spacing on a diameter of 16 mm, around a 5 mm-diameter central post. The axial, center-to-center separation of the arrays is 27 mm, such that

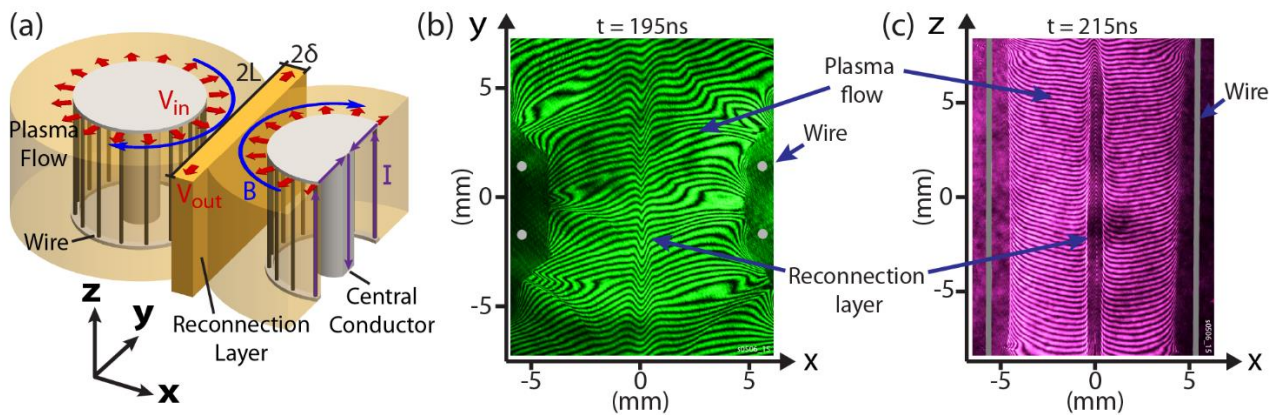

FIG.1 (a) Schematic diagram of the experimental setup (with cut-away of the right wire array): current is applied in parallel to two inverse wire arrays, producing magnetized plasma flows which collide to create a reconnection layer. The directions of the current (purple), plasma flows (red) and the embedded magnetic fields (blue) are shown. (b-c) Raw interferometry images of the interaction region following the formation of the reconnection layer, showing (b) the xy-plane and (c) the xz-plane, as defined by the Cartesian coordinate system in (a). The positions of the wires are indicated.

the minimum gap between the wires of the two arrays is 11 mm. The arrays are driven with the same polarity, such that when the advected magnetic fields meet they are orientated in opposite directions, and their interaction leads to the formation of a reconnection layer at the mid-plane.

The reconnection layer is diagnosed using a suite of complementary plasma diagnostics. These diagnostics can be fielded simultaneously, allowing the dynamics of the interaction and the localized plasma parameters of the system to be determined with a high degree of spatial and temporal resolution. Due to the highly reproducible nature of the plasma formation and evolution in this setup, the diagnostics can be used to acquire data from different times throughout the development of the interaction across multiple experiments. The details of the diagnostic setup are summarized as follows.

To obtain a qualitative overview of the morphology and structural evolution of the system over the course of a single experiment, the dynamics of the interaction are captured using a high-speed, multi-frame, optical camera (Invisible Vision U2V1224: 12 frames, 5 ns exposure, tuneable interframe time $\Delta t \geq 5$ ns, with a 600 nm low-pass filter to block light at laser diagnostic wavelengths). With reference to the Cartesian coordinate system defined in Fig. 1(a), the camera images the self-emission from the plasma along the z-direction, thus producing images of the xy-plane of the interaction region as demonstrated in Fig. 2.

Several laser-based diagnostics are employed to measure the quantitative features of the plasma structure. Measurements of the (line-integrated) electron density distribution of the interaction region are made using Mach-Zehnder interferometry imaging[31]. Interferograms of the xy-plane (Fig. 1(b)) are obtained by probing along the z-direction (parallel to the axes of the arrays), using the 2nd (532 nm) and 3rd (355 nm) harmonics of a pulsed Nd:YAG laser (EKSPLA SL321P, 0.5 ns, 500 mJ). Both the 532 nm and 355 nm channels use the same probe path, but have a time offset to provide two interferograms separated by 20 ns. A separate interferometer probes the plasma perpendicularly, along the y-direction, producing interferograms of the xz-plane (Fig. 1(c)), using an independent, 1053 nm, 1 ns, 5 J probe beam. The interferograms are recorded by Canon 350D and 500D DSLR cameras, with the shutters held open for the duration of the experiments, such that the time resolution is set by the pulse duration of the probe laser beams. The interferograms are processed to produce maps of electron line density ($\int n_e dl$), using the analysis procedure described in Refs. 31,33.

The magnetic field distribution is measured using a Faraday rotation polarimetry diagnostic[34]. The polarimetry is performed in the y-direction using the same 1053 nm probe beam as the xz-interferometer. The line-averaged field strength along the probe direction is calculated from the rotation of the linear polarization of the probe beam. The diagnostic consists of two channels, with oppositely offset linear polarizers, set at 3° either side of extinction, and two identical Atik 383L+ CCD cameras. For each channel, the spatial distribution of rotation angle is determined via the change in intensity recorded on the CCDs due to the rotation of polarization, either towards or away from extinction. The combination of the two polarimetry channels allows the optical self-emission from the plasma to be removed, reducing systematic errors in determining the polarization of the laser beam. Further details of the diagnostic are described in Ref. 31.

An optical Thomson scattering (TS) diagnostic system ($\lambda$=532 nm, 5 ns FWHM, 3 J) records the ion feature of the collective TS spectra from within the interaction region[35]. A focussed laser beam is passed through the xy-plane at the mid-height (z=0 mm) of the arrays, with a waist diameter of $\sim$200 µm throughout the entire range of interest of the plasma. The scattered light is collected using single lens systems to image the path of the laser beam onto the input of fiber-optic bundles. The bundles contain 14 individual, 100 µm-diameter fibers, each collecting light from a separate scattering volume on the beam path. In the majority of experiments two independent imaging systems were used, observing the TS light from matching spatial positions, but along different scattering directions in the xy-plane. In other experiments a single imaging system was used, collecting scattered light in the out-of-plane, z-direction. Further details of these scattering geometries are provided in Sec. IIIC. The coordinates of the scattering volumes are identified within the xy-plane of interferometry images, to a precision of $\leq$200 µm, using the procedure described in Ref. 31, allowing the positions of the TS measurements within the plasma structure to be determined. The output from the fiber-optic bundles is recorded using an imaging spectrometer (ANDOR SR-500i-A, with gated ANDOR iStar ICCD camera). The spectral resolution of 0.25 Å is set by the combination of the $\sim$50 µm spectrometer slit width and 2400 lines/mm grating, and the temporal resolution is set by the 4 ns gate time. The Doppler shift of the TS spectra allow the components of the bulk plasma flow velocity to be calculated along each of the scattering vectors defined by the observation directions (see

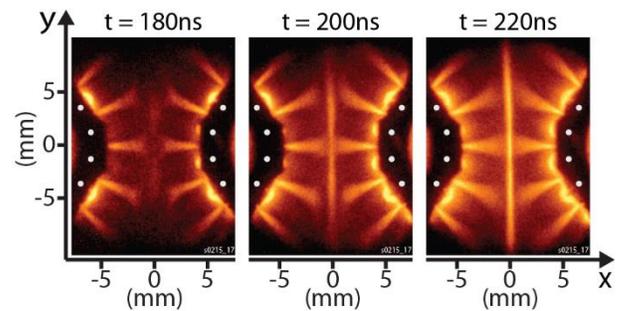

FIG.2 Time series showing optical self-emission images (false-color) of the plasma in the xy-plane, from a single experiment. White dots indicate the wire positions. Video available online covering the time interval t=160-380 ns (Multimedia view).

Sec. IIIC and Fig. 5). Additionally, by fitting theoretical form-factors to the profiles of the spectra[31,35], and utilizing the electron density values obtained from interferometry measurements, the local ion temperature ($T_i$) and the product of the average ionization and electron temperature ($\bar{Z}T_e$) of the plasma can be extracted. A non-local thermodynamic equilibrium (nLTE) model[36] is used to decompose $\bar{Z}T_e$ into self-consistent values of $\bar{Z}$ and $T_e$.

## III. EXPERIMENTAL RESULTS

### A. Formation of the reconnection layer

The collision of the magnetized flows in the mid-plane between the two wire arrays leads to the formation of the reconnection layer. This is seen in the optical, self-emission image time-series presented in Fig. 2 (video available online). The images show that the layer becomes detectable with this diagnostic during the time interval t=160-180 ns after the start of the ~500ns duration current pulse (the current start is used as the reference for all times quoted in this paper). This delay between current start and the formation of the layer originates from the combination of the "dwell" time for the first ablated plasma to be formed at the wires, and the time-of-flight of the plasma to reach the mid-plane. Measurements from previous wire array experiments on MAGPIE have typically shown a dwell time of ~50 ns[37], and thus a plasma flow velocity into the reconnection region can be estimated as $V_{in} \geq 5.5$ mm$/(120 \pm 10$ ns$) \approx 40 - 50$ km/s. Following the reconnection layer formation, the intensity of self-emission in the layer increases. The observable length of the reconnection layer also increases, rapidly expanding outwards from the center of the images along the y-direction, reaching the bounds of the field-of-view (y = $\pm 11.5$ mm) by t=240 ns. Throughout the experiments the layer appears notably straight and maintains an approximately constant thickness until late in the experiments (t>300 ns) when the drive current has passed its peak. Subsequently, the ram pressure of the flows is expected to start decreasing, which is consistent with the broadening of the layer as it starts expanding against the upstream flow at late time.

Interferograms of the interaction region are obtained along both the y- and z-directions, as demonstrated by the examples in Figs. 1(b) and 1(c) respectively. These

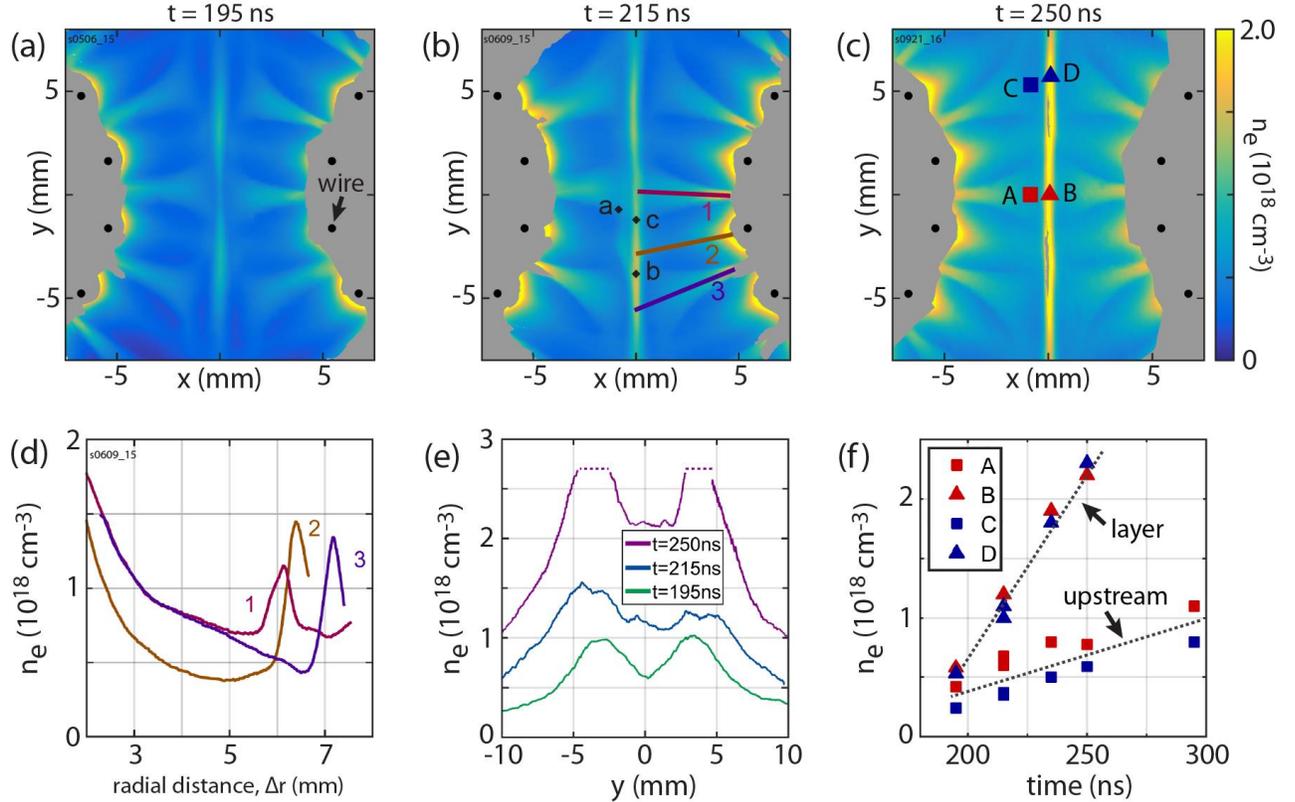

FIG.3 (a-c) Electron density ($n_e$) distributions calculated from interferograms of the xy-plane of the interaction region, at three times in three separate experiments. Regions where the interferometry fringes could not be traced are masked in gray. The positions marked a-c in (b) correspond to the coordinates of TS measurements discussed in Sec. IIIA, whose spectra are presented in Fig. 6. (d) Electron density profiles of the plasma flow along the radial paths indicated in (b). (e) Electron density profiles along the length of the reconnection layer from the density maps in (a)-(c) (Data denoted by dashed lines are lower-limits, as $n_e$ could not be directly measured at these positions due to sharp local density gradients obscuring the interferometry fringes.). (f) Electron density vs. time at the positions marked A-D in (c).

interferograms reveal the line-integrated electron density distribution of the plasma by the localized bending of the initially straight (horizontal) interference fringes, which is strongest at the mid-distance between the arrays. The displacement of the fringes is proportional to the integral of the electron density along the line-of-sight through the plasma ($\int n_e dl$). Thus, the raw interferograms show that the reconnection layer has a greater electron density than the incoming plasma flows, and is uniform in the axial, z-direction over the height of the arrays. In agreement with the self-emission images, the layer is first observed in the interferometry data at t≈180 ns, when the fringe shift reaches approximately half a fringe spacing, equivalent to $\int n_e dl = 2 \times 10^{17} cm^{-2}$.

The electron density distribution in the reconnection plane is measured using the xy-plane interferometer. The raw interferograms, similar to that shown in Fig. 1(b), are processed into maps of electron line density using the procedure described in Refs. 31,33. These are further converted to electron density ($n_e$) by dividing by the axial height of the arrays (Δz=16 mm), utilizing the uniformity of the plasma structure in this direction (Fig. 1(c)). Figs. 3(a)-(c) present typical electron density distribution maps obtained at different times in the experiments, demonstrating the temporal evolution of the layer. The radially diverging plasma flows propagating from the arrays can be seen on the left and right-hand sides of these images. The ablated plasma density is modulated azimuthally about the arrays due to the finite number of wires, and the regions of higher density correlate with the features observed in the self-emission images.

The flow structure consists of ablation from each of the wires, as well as regions of enhanced density between the wires, formed by the collision of plasma expanding from the adjacent wires. These collision regions are bound by oblique shock fronts, analogous to the structure observed in the interior of imploding aluminum wire arrays[33], and are indicative of the supersonic velocity of the flows. The density within the shock-bound regions is greater than that of the ambient flow. This is evident from a comparison of radial profiles at varying azimuthal positions, e.g. profiles 1-3 in Fig. 3(d), corresponding to the line-outs marked in Fig. 3(b). These profiles also show that at each azimuthal position the flow density upstream of the layer decreases with radial distance (Δr) from the array. This is due to the combination of the time-of-flight of the flow (with lower ablation density produced earlier in time when the current is smaller) and the cylindrical divergence of the flow geometry. The plasma density rises at the position where the flow meets the boundary of the reconnection layer, just ahead of the mid-plane. The precise overlap of the profiles 1 and 3 for the range Δr<5 mm, demonstrates not only that the flows inside the shock-bound regions are identical, but that the upstream flow at distances of |x|>1 mm from the layer is not disturbed by the existence of the layer, which is consistent with the supersonic nature of the ablation flows. The flows are also not expected to penetrate through the layer to the opposing side, as the mean free paths of the plasma particles are significantly shorter than the layer thickness (see Sec. IV for more details). Despite the upstream equivalence of the shock-bound flow regions shown in Fig. 3(d), the density of the outer stream (3) at the boundary of the layer is lower, due to the longer path length to the layer and hence greater cylindrical divergence undergone. Thus, the greatest flux of plasma into the layer occurs along the central direction y=0 mm.

Inside the layer, the electron density is significantly larger than in the upstream flow: e.g. at t=215 ns (Fig. 3(d)) the factor of increase is in the range of 1.5-3. The maximum density, however, is not located at the central position of the layer (x,y)=0 mm, despite it receiving the highest density from the upstream flow. This is demonstrated by the electron density profiles presented in Fig. 3(e), which show plots along the layer from the distributions in Figs. 3(a)-(c). The profiles reveal peaks of density located at symmetric positions either side of the layer center, with the separation of these peaks increasing with time. The rate of displacement of the peaks is consistent with the flow of material outwards along the layer at a velocity in the range of 30-50 km/s. This agrees well with direct measurements of the layer outflow velocity ($V_y$) using Thomson scattering, which are presented in Sec. IIIC.

In accordance with the optical, multi-frame images (Fig. 2), the electron density distributions show that for t < 300 ns, the layer maintains an approximately uniform thickness along its entire length (FWHM: $2\delta = 0.6$ mm), while the layer expands in the perpendicular y-direction. In the image of Fig. 3(a), taken at t=195 ns, the layer displays a length of Δy=13 mm, which rapidly increases to 15 mm by t=215 ns (Fig. 3(b)), and later extends beyond the bounds of the field-of-view, i.e. Δy≥16 mm (Fig. 3(c)). The outflow of material along the layer plays a dominant role in this expansion process. Evidence of this can be seen in Fig. 3(f), where the electron density is plotted as a function of time for positions upstream and inside the layer. The data series are labelled A-D corresponding to the positions of the measurements indicated in Fig. 3(c). At the center of the layer (position B) the density is comparable to that in the outflow (position D). This is despite a much higher upstream density flowing into the center of the layer (position A), than in the corresponding, off-center upstream flow (position C). Thus, the outflow along the layer must make up a significant contribution to the density at position D to offset this difference in the inflows.

**B. Measurements of the magnetic field distribution**

The distribution of the magnetic field in the xz-plane was measured using the simultaneous interferometry and polarimetry diagnostics. Figs. 4(a)-(c) present data obtained with these diagnostics at a time of t=195 ns, shortly after the layer formation, and at the same time as the xy-plane density map shown in Fig. 3(a). These maps depict the xz-plane of the interaction region, with the flows moving horizontally inwards from the arrays positioned at the left and right-hand edges of the field of view. The electron line-integrated density map in Fig. 4(a) shows the density increase in the reconnection layer in comparison to the flows, with the thickness $2\delta$ of the layer matching that observed in the xy-plane. The spatial variation of the rotation angle of the linear polarization of the probe beam is shown in Fig. 4(b). The rotation angle is sensitive to the $B_y$-component of the magnetic field, parallel to the direction of the probing beam through the plasma. The rotation is symmetric with respect to the midplane (x=0 mm) of the interaction region, with equal and oppositely directed rotation angles of $\alpha = \pm 0.2°$ measured in the plasma on either side of the layer. This is consistent with the expected magnetic field geometry of the experimental setup (Fig. 1(a)) of oppositely directed fields embedded in the flows from each array.

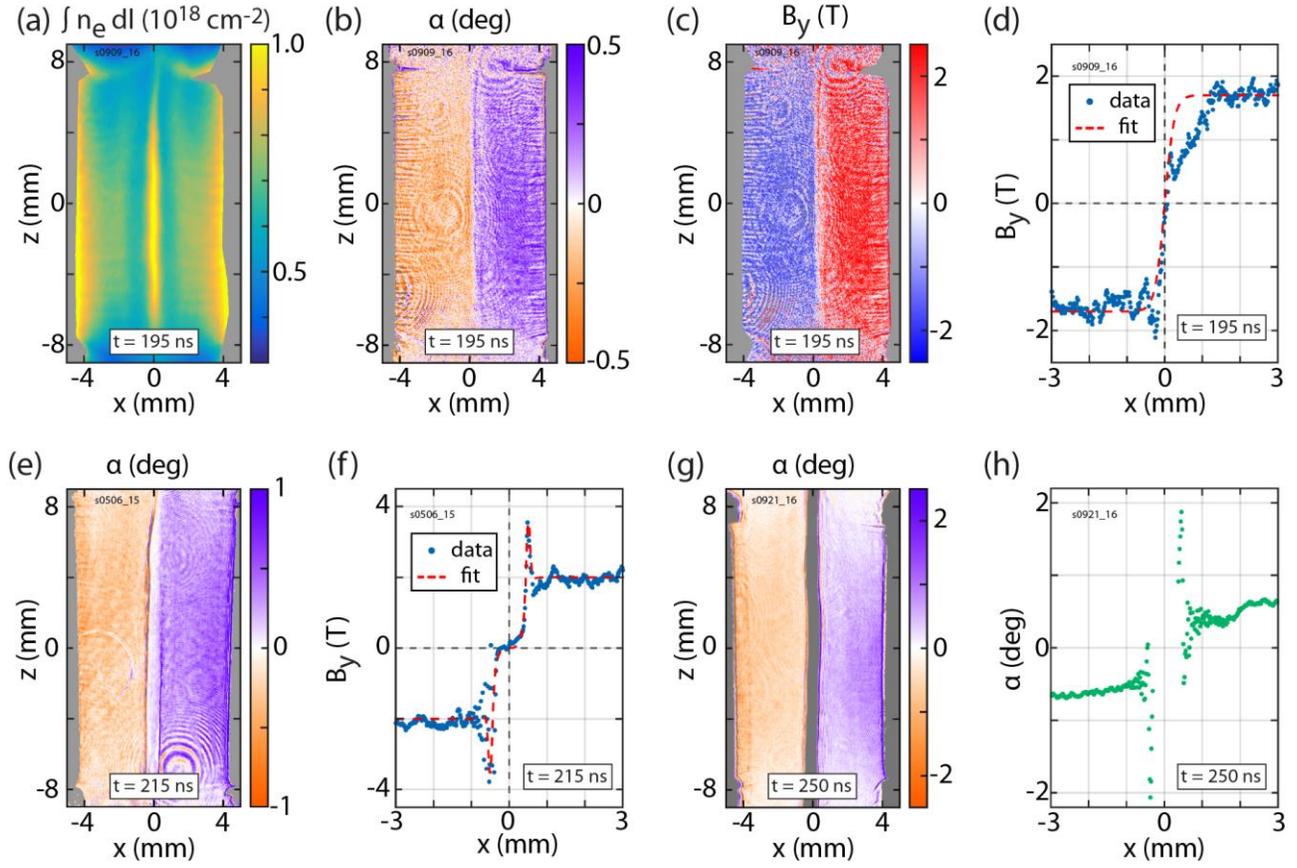

FIG.4 Faraday rotation polarimetry data. (a) Line-integrated electron density distribution of the xz-plane at t=195 ns. Regions where the probe beam was obscured are masked in gray. (b) Angular rotation distribution of the probe beam at t=195 ns. Concentric circular features are artefacts from diffraction of light around dust spots on the optics of the imaging system. (c) Magnetic field distribution calculated from the combination of data in (a) and (b). (d) Horizontal profile of the magnetic field in (c), averaged vertically over the range z=-3 to 3 mm. (e) Rotation distribution at t=215 ns, showing enhancements of the Faraday rotation angle at the boundaries of the layer. (f) The accompanying magnetic field profile for (e). (g) Rotation distribution at t=250 ns, showing further increasing field enhancement at the layer boundaries. (h) Rotation profile for (g). Data points were not obtained inside the layer at this time as the laser beam could no longer probe through the layer.

The Faraday rotation angle is determined by both the magnetic field and electron density of the plasma. The average $B_y$-component of magnetic field can be found by dividing the rotation angle by the electron line density, using the formula[34]

$$B_y(x,z) = \frac{8\pi^2 \varepsilon_0 m_e^2 c^3}{e^3 \lambda^2} \frac{\alpha(x,z)}{\int [n_e(x,y,z)dy]}. \quad (1)$$

The resulting magnetic field distribution is displayed in Fig. 4(c). The distribution exhibits notable uniformity in the z-direction, despite the presence of some noise on small spatial scales. To suppress the noise, a horizontal profile is taken through the field map, averaged vertically over the interval z=−3 to 3 mm, producing the plot shown in Fig. 4(d). The field profile shows that upstream of the reconnection layer the magnetic field has an approximately constant strength, with $B_y=\pm 1.7$ T measured on either side of the layer. Inside the layer this field drops steeply, passing through 0 at the mid-point between the arrays. This profile can be well approximated by a "Harris-sheet" form ($B_y = B_0\tanh[x/\delta]$, red dashed line in Fig. 4(d)), typically used to describe magnetic reconnection current sheets, and the underlying current density distribution $(j_z = -1/\mu_0\, \partial B_y/\partial x)$ for this fitted magnetic field profile corresponds to a peak current density of 0.5 MA/cm².

At later times in the experiments the increasing density gradients inside the reconnection layer are large enough to refract the probe laser beam beyond the acceptance angle of the imaging system. Consequently, soon after the formation of the layer, measurements of the magnetic field deep inside the layer become unreliable, but the field can still be measured in the flows upstream and at the boundaries of the layer. Fig. 4(e) displays a map of the Faraday rotation angle at a time of t=215 ns (corresponding to the time of Fig. 3(b)). At this time the Faraday rotation angles in the upstream flow have increased to $\alpha = \pm 0.3°$, indicating greater embedded magnetic field and plasma density. Additionally, even stronger symmetric rotation angles of $\alpha = \pm 1.0°$ are observed over narrow intervals of $\Delta x \sim 0.1$ mm at the boundaries (x=$\pm 0.5$ mm) of the layer, which are in the same directions as the rotation in the adjoining upstream flows. These sharp features at the layer boundaries are not simultaneously observed by the interferometry diagnostic, and so it can be concluded that they signify considerable enhancements of the magnetic field at these locations. The magnetic field profile calculated from this data (Fig. 4(f)) shows that at t=215 ns the upstream flows possess field strengths of $B_y = \pm 2$ T and the enhanced fields at the edges of the layer are $\pm 4$ T.

Polarimetry data obtained at subsequent times in the experiments show that the magnetic field brought by the upstream flow continues to increase, and that the field enhancements at the boundaries of the layer persist, also with an increasing strength. This is demonstrated by the Faraday rotation distribution and accompanying profile in Figs. 4(g) and 4(h) respectively, which are taken at t=250 ns and concurrent with Fig. 3(c). The Faraday angles in the upstream flows ($\alpha = \pm 0.5°$) correspond to magnetic fields of $B_y = \pm 4$ T, while the rotations of $\pm 2°$ at the layer boundaries correspond to $\pm 8$ T.

### C. Spatially resolved Thomson scattering measurements of flow velocities and plasma temperature

- **Measurements in the xy-plane**

The multipoint Thomson scattering (TS) diagnostic operated simultaneously with the interferometry and polarimetry measurements, allowing detailed measurements of the local flow velocities and thermal properties of the plasma. The TS measurements presented in this sub-section were obtained from the geometry illustrated in Fig. 5(a), in which the probing laser beam passed through the reconnection layer at an angle of 22.5° to the y-axis. The scattered light was observed from two opposing directions, corresponding to scattering angles of 45° and 135°. In both directions the light was collected from 14 matching, equally-spaced positions along the linear path of the beam, which

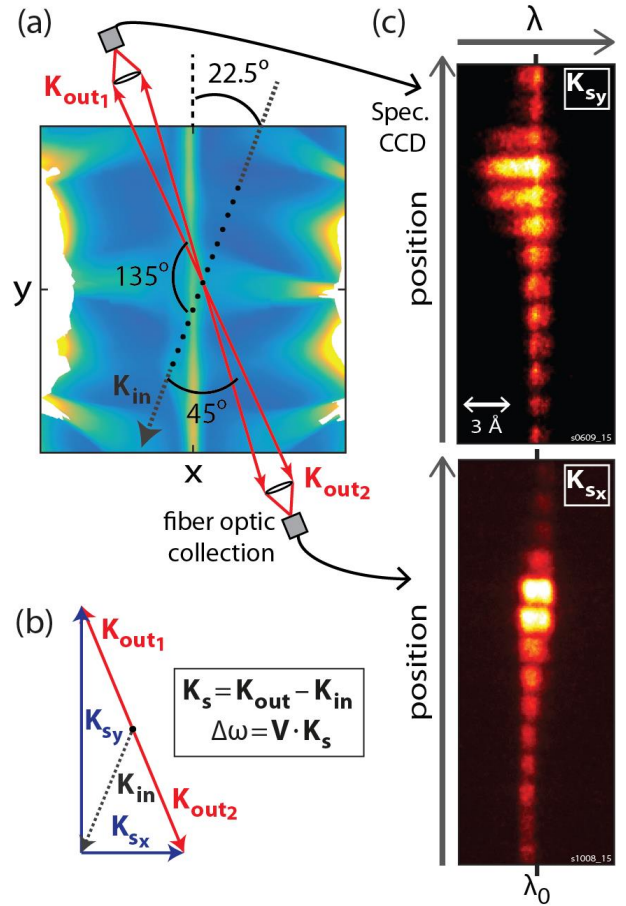

FIG.5 (a) Schematic diagram of the Thomson scattering geometry used to make independent measurements of $V_x$ and $V_y$. TS light is observed from 14 positions along the path of the probe beam from two opposing directions. The 14 scattering volumes are imaged onto the input of individual optical fibers, coupled to an imaging spectrometer. (b) Geometry of the probe laser, observation and scattering K-vectors. Equations show the vector relations and dependence of the Doppler shift ($\Delta\omega$) of the spectra on the local flow velocity and scattering vector. (c) Examples of raw TS spectra from the imaging spectrometer CCD. The vertical axis of the spectrometer corresponds to the position along the probe beam.

was achieved by imaging the beam path onto the inputs of fiber-optic bundles. The imaging systems had a magnification of 0.8, giving collection volumes 125 µm in diameter, spaced by 0.3mm. The geometry of the input laser beam and the observation directions defines scattering vectors ($\mathbf{K_S} = \mathbf{K_{out}} - \mathbf{K_{in}}$), as depicted in Fig. 5(b). The detection of the Doppler shifts ($\Delta\omega = \mathbf{V_{flow}} \cdot \mathbf{K_S}$) of the TS spectra provides measurements of the components of the bulk velocity of the plasma, along the direction of the corresponding scattering vectors. The geometry of this TS setup was selected to obtain separate measurements of the orthogonal components $V_x$ and $V_y$ of the velocity, which together reveal the speed and direction of the plasma flow within the xy-plane.

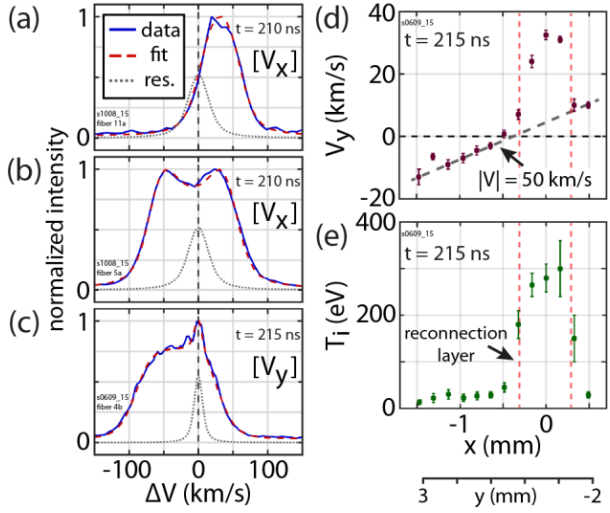

FIG.6 (a-c) Fitted TS spectra for the three spatial positions in the interaction region marked in Fig. 3(b). The dashed profile (res.) indicates the spectrometer response function. (d-e) Profiles of $V_y$ and $T_i$ measured in a single experiment for scattering volumes along the TS beam when passing through position c in Fig. 3(b). The data is plotted as a function of function of x-position, with the additional scale below indicating the corresponding y-positions of the volumes.

Fig. 5(c) shows examples of raw TS spectra from the fiber-optic bundles of each of the two observation directions. The spectrometer CCD images display the discrete spatial positions through the plasma (i.e. fibers) on the vertical axes, against the spectrum of the scattered light along the horizontal axes. The spectral shape of the TS signal is fitted with a theoretical spectrum to infer the temperature of both the electron and ion populations of the plasma. The TS spectra are processed by integrating vertically across the CCD pixels for each individual fiber, and fitting is performed using the non-relativistic, Maxwellian spectral density function $S(\omega,\mathbf{K})$[35]. As part of this fitting procedure, the theoretical spectrum is convolved with the response function of the spectrometer, which is found from the observed broadening on the unshifted laser wavelength, recorded before the experiment.

Figs. 6(a)-(c) present examples of fitted TS spectra, displayed with their horizontal axes converted from wavelength to velocity ($V = [2\pi c/(\lambda_0 + \Delta\lambda) - \omega_0]/K_S$). The spectra are from 3 key spatial positions in the reconnection xy-plane, taken around the time of Fig. 3(b) (t=210-215 ns). The black dots marked a-c on the density map of Fig. 3(b) denote the coordinates of the collection volumes for these spectra. The $V_x$-sensitive spectrum of Fig. 6(a) was obtained in the flow upstream of the layer at (x,y)=(-1,-1) mm, and shows that the flow here approaches the layer with a velocity component perpendicular to the layer of $V_x$=40 km/s. Inside the layer boundary (x<δ), the $V_x$ component of the flow is found to rapidly fall to 0 (e.g. Fig. 6(b)), however, for coordinates y≠0 mm the plasma inside the layer acquires significant outflow motion in the y-direction (e.g. Fig. 6(c)).

The presence of a significant outflow along the layer is clearly seen from a comparison of the $V_y$ velocity components measured in the layer and in the upstream plasma. Fig. 6(d) contains a full set of measured $V_y$ velocity components from a single experiment where the laser crossed the layer at y=1.7 mm. The dataset covers the range (x,y)=(-1.5,3) mm to (0.5,-2) mm. The profile shows that non-zero y-velocities are present outside of the layer, but these follow a strict trend (indicated by the dashed line in Fig. 6(d)), consistent with the picture of cylindrically divergent flow emanating from the arrays at a constant radial speed of $|\mathbf{V}| = 50$ km/s. Inside the layer, however, there is a significant deviation from the linear profile of $V_y$ velocity components in the upstream plasma (dashed line), which is greatest at x=0 mm. The FWHM of the region with high outflow velocities closely matches the width of the layer measured in the electron density structure. Similar measurements, performed in experiments where the TS probe beam crossed the layer at y=1.0 mm and 3.7 mm, yielded outflows of $V_y$=30 km/s and 60 km/s respectively, indicating an outward acceleration of material along the layer. The results are consistent with the inferred outflow velocities across multiple frames of the interferometry data, described in Sec. IIIA.

In addition to the formation of fast plasma outflows along the layer, strong ion heating is observed inside the layer during the early part of the experiments (t=195-225 ns). Local temperatures obtained from spectral fits for both scattering directions show good agreement, strongly suggesting that the shape of spectra from the layer is determined by thermal motion (i.e. temperature), and not by variations of the bulk flow velocity within the scattering volumes. It is noted however that such measurements cannot fully exclude contributions from small-scale turbulent motions of the plasma, but this would require motions on spatial scales smaller than the size of the collection volumes (i.e. <<125 µm). The widths of the spectra measured inside the layer at t=215 ns were significantly broader than those upstream. The upstream plasma was found to be cold ($T_i$=22±10 eV, $T_e$<20 eV), while inside the layer the ion temperature reached $T_i$∼300 eV (Fig. 6(e), corresponding to the same dataset as Fig. 6(d)).

The electron temperature in the layer is best determined from the data obtained at the θ=45° observation angle due to a higher value of the scattering parameter[35]

$$\alpha = \frac{1}{K_s\lambda_D} \propto \frac{1}{\sin(\theta/2)}. \qquad (2)$$

For α>1 spectra display ion acoustic peaks (e.g. Fig. 6(b)), whose separation is sensitive to the product of the electron temperature and average ionization, $\bar{Z}T_e$. In the case of Fig. 6(b) the data were best fitted with a value of $\bar{Z}T_e = 320 \pm 20$ eV, corresponding to values of $\bar{Z} = 7.3$ and $T_e = 43$ eV in the nLTE model. It is emphasized that in all TS data collected during the time interval t=195-225 ns, the measured ion temperature in the reconnection layer was found to significantly exceed the electron temperature inside the layer, and that $T_i \approx \bar{Z} T_e$.

- **Measurements in the yz-plane**

The TS measurements detailed thus far were made with the probe laser crossing the layer in the xy-plane, in the geometry of Fig. 5(a), and the local parameters of the reconnection layer at different positions along the layer were obtained by performing multiple experiments and varying the y-coordinate of the crossing. This method however restricts the extent of the layer which can be studied, due to the probe path being obstructed by the array hardware at large y-crossing values. To overcome this limitation an alternate TS geometry was employed, where the trajectory of the probe beam was along the y-direction, at the mid-height of the arrays (z=0 mm), and the scattered light was observed at 90° in the out-of-plane z-direction, using a single fiber-optic bundle. This scattering geometry defined an alternate $K_S$ vector, directed at 45° to $\hat{y}$ and $\hat{z}$, such that the Doppler shift was equally sensitive to $V_y$ and $V_z$ ($\Delta\omega = [K_S/\sqrt{2}][V_y + V_z]$).

Fig. 7(a) shows the positions of 14, 400 μm-diameter scattering volumes along the reconnection layer in an experiment using this TS geometry. The velocities measured in this experiment are shown in Fig. 7(b), with the data points in red (circles) calculated directly from the observed Doppler shifts of the scattered spectra, and thus giving the velocity component in the direction of the scattering vector. In agreement with measurements performed in the xy-geometry, they show the presence of large velocities inside the layer. However, whilst the shape of the velocity profile is symmetric about y=0 mm, there is a systematic, positive offset in the velocities. Since measurements performed in the layer in the xy-plane demonstrated that $V_y$=0 km/s at y=0 mm, it can be concluded that the Doppler shift at this position must be attributed to flow in the positive z-direction, equal to ∼50 km/s. Under the assumption that this $V_z$ component is constant along the layer, and equal to the value of $V_z$(x=0,y=0), the $V_y$ component along the layer can be calculated as

$$V_y(y) = \frac{V_S(y)}{\sin(45°)} - V_z(0,0). \quad (3)$$

These values are plotted in Fig. 7(b) as the blue data points (squares). It is seen that this $V_y(y)$ profile is symmetric about the central position of the layer y=0 mm, and that the outward flow of plasma reaches velocities of $V_y$=±100 km/s by y=±10 mm. It is important to note that measurements were also performed with the probe beam propagating in the y-direction, similar to Fig. 7(a), but with the beam path just outside of the layer, along x=2δ=0.6 mm. In this case there was little evidence of either vertical motion, or motion parallel to the layer beyond the standard cylindrical divergence of the flow, previously discussed in conjunction with Fig. 6(d). Thus, it can be concluded that the strong outflow of plasma from the layer, demonstrated in Fig. 7(b), is due to plasma being accelerated inside the layer.

- **Temperature measurements at late time**

During the later experimental times of t≥240 ns, the shape of TS spectra obtained at all spatial positions along the layer were significantly different to those described thus far. An example late-time spectrum, recorded in the xy-plane using the $V_y$ sensitive scattering angle at t=295 ns, is shown in Fig. 8(a). The spectrum shows a narrower width than observed for earlier times, indicating a lower ion temperature, and displays more pronounced ion acoustic peaks, corresponding to a condition of $T_i \ll \bar{Z} T_e$. This spectral shape allows the temperature parameters to be fit to a high degree of precision, and in combination with the co-constrained, simultaneous $V_x$-sensitive spectrum, the fitting yields $T_i = 33 \pm 5$ eV and $\bar{Z} T_e = 135 \pm 6$ eV, corresponding to values of $T_e = 25$ eV and $\bar{Z} = 5.4$ in the nLTE model. The temporal evolution of the temperature parameters measured inside the layer is summarized in Figs. 8(b) and 8(c). The first of these plots demonstrates that during the early-time interval t=190-240 ns there is a continuous cooling of the ions in the layer, with a characteristic timescale of $T_i/(dT_i/dt)$ ∼35 ns. After this, for the late times of t≥240 ns, the ions in the layer maintain a temperature of around 30 eV. In contrast, the second plot shows that there is very little change in the measured ionization electron temperature product, which corresponds to an electron temperature varying across the range $T_e$ =40→30 eV in the transition from early to late times. It

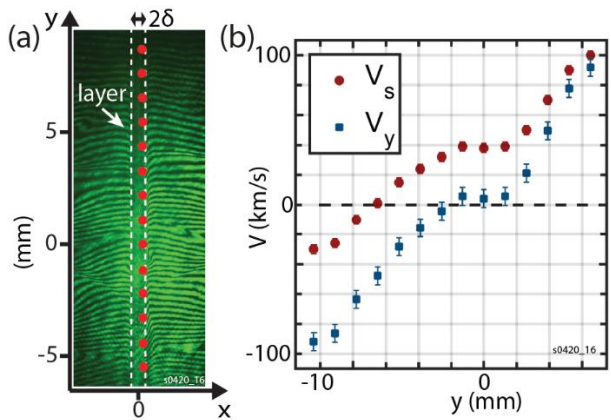

FIG.7 (a) Positions of TS volumes in an experiment with the probe beam passing along the reconnection layer. The red dots show the size of the scattering volumes to scale. The white dashed lines indicate the boundaries of the reconnection layer, as observed in the interferogram which was obtained simultaneously. (b) Velocity measurements from this experiment at the positions depicted in (a). The red dots show the velocity measured in the direction of the $K_s$ vector, which was aligned at 45° to $\hat{y}$ and $\hat{z}$, and therefore equally sensitive to $V_y$ and $V_z$. The blue squares show $V_y$ calculated from the measurements using Eq. (3).

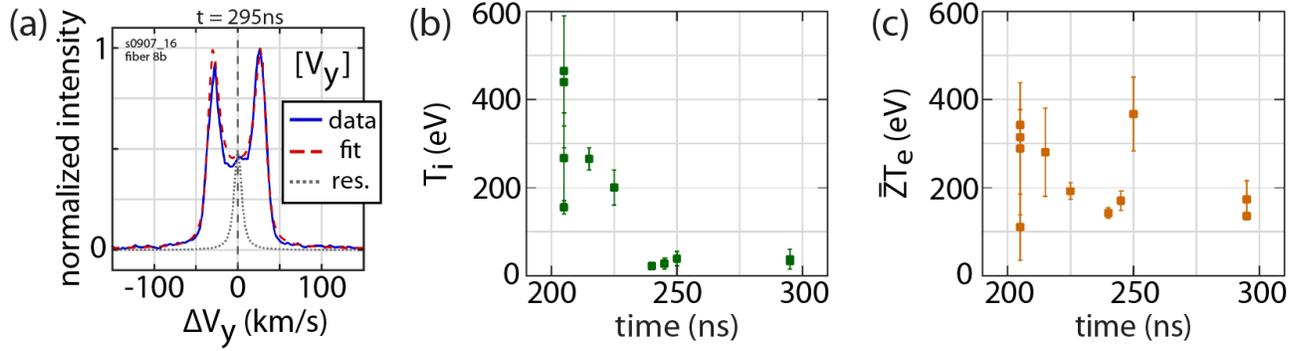

FIG.8 (a) Example of a TS spectrum typical for t≥240 ns, showing narrow and pronounced ion-acoustic peaks recorded at the center of the layer (x,y)=0 mm, using the $V_y$-sensitive geometry depicted in Fig. 5. (b-c) $T_i$ and $\bar{Z}T_e$ measured inside the reconnection layer at different times. Each data-point corresponds to the most centrally-located measurement inside the layer from an individual experiment.

is also important to note that at all times Thomson scattering measurements showed a $V_y$ profile consistent with that in Fig. 7(b).

## IV. SUMMARY AND DISCUSSION

The experiments reported in this paper utilize counter-streaming, supersonic, magnetized plasma flows, with anti-parallel magnetic fields, to produce a reconnection layer in which the magnetic flux is annihilated. The flow parameters of the setup provide the boundary conditions for the magnetic reconnection process, which in this case is "strongly driven" due to the high ratio of ram to magnetic pressure in the inflows (characterized by a high dynamic Beta parameter, $\beta_{dyn} = \rho V_{flow}^2/[B^2/2\mu_0] \sim 10$). Consequently, the velocity of the inflows is super-Alfvénic ($M_A \sim 2$).

A notable feature of these pulsed power driven experiments is the long duration of the reconnection layer. In contrast to the more transient reconnection phenomenon occurring in laser-driven reconnection experiments[14–19], the reconnection layer appears to be in a stable and approximately steady-state, maintained by a continuous inflow of plasma with embedded magnetic field for a timescale >100 ns. This is many times greater than the hydrodynamic time-scale of the system, which can be estimated as the time taken for the inflow to cross a spatial scale equal to the layer thickness, i.e. $\delta/V_x \sim 5$ ns. The measurements presented in this paper focus on the characterization of the reconnection layer plasma parameters, which reveal in detail the structure of the reconnection layer and its evolution over the observed timescale. An overview of the measured plasma parameters at two times in these experiments, representative of the early and late time properties of the layer, is given in Table I and the main observations of the study are discussed below.

### A. Structural features of the reconnection layer

The geometry of the experiments is quasi-two-dimensional: the xy-plane of the setup defines the (reconnection) plane in which the reconnecting magnetic field lines lie (Figs. 1(b) and 2), and in the perpendicular xz-plane there is a good, linear symmetry (Fig. 1(c)). The setup does not contain any guide field. Measurements of the magnetic field and density distributions of the plasma show that the magnetic flux advected by the inflows is annihilated inside the layer, and that there is an accompanying increase in the plasma density inside the layer (Figs. 3 and 4(a)-(d)). Following the initial formation of the layer, the magnetic field distribution is found to closely resemble a Harris sheet profile (Fig. 4(d)), and the half-thickness $\delta$ of the sheet matches that of the density rise in the layer. At subsequent times, strong and narrow enhancements in the local magnetic field strength develop in narrow intervals ($\Delta x \lesssim 0.1$ mm) at the boundaries of the layer (Figs. 4(e) and 4(f)), consistent with the pile-up of magnetic flux, and these features increase in prominence over time (Figs. 4(g) and 4(h)). It is interesting that the half-thickness of the layer is equal to the ion skin depth of the plasma $d_i = c/\omega_{pi}$ calculated from the layer plasma parameters. This suggests that two fluid physics, such as the Hall effect[10–13], may play an important role in this system, as the ions decouple from the electrons on the spatial scale of the flux pile-up. The mean free paths of both electrons and ions in the plasma are much shorter than the spatial scales of all observed features of the layer structure ($\lambda_{ii} \sim 10^{-2}$ mm, $\lambda_{ei} \sim 10^{-3}$ mm), so the plasma is strongly collisional.

An analysis of the mass flowing into and out of the central region of the reconnection layer, with the bounds $|y| < 0.8$ mm and $|x| < \delta$, reveals that the rate of plasma inflow to the layer $(\Delta y V_x n_{i,in})$ is approximately a factor of 2 greater than the outflow rate $(2\delta V_y n_{i,out})$. The accumulation of material in the reconnection layer is indeed seen in the increasing electron density in Fig. 3(e). Conversion of this measured $n_e(t)$ to ion density, using local TS measurements of the up- and downstream ionization states of the plasma, shows a very good agreement with the flux estimate. Despite the changes in the material density and plasma temperature in the layer (Fig. 8(b) and 8(c)), the thickness of the layer is approximately constant throughout the experiments. An

TABLE I. Plasma parameters of the inflow and reconnection layer at early and late times in the experiments.

| Parameter | Symbol | t = 215ns Inflow | t = 215ns Layer | t = 250ns Inflow | t = 250ns Layer |
|---|---|---|---|---|---|
| Electron temperature (eV) | $T_e$ | 15 | 40 | 15 | 30 |
| Ion temperature (eV) | $T_i$ | 20 | 300 | 20 | 30 |
| Ionization | $\bar{Z}$ | 3.5 | 7 | 3.5 | 5.7 |
| Electron density (cm$^{-3}$) | $n_e$ | $5\times10^{17}$ | $1.3\times10^{18}$ | $8\times10^{17}$ | $2.3\times10^{18}$ |
| Ion density (cm$^{-3}$) | $n_i$ | $1.4\times10^{17}$ | $1.9\times10^{17}$ | $2.3\times10^{17}$ | $4\times10^{17}$ |
| Magnetic field (T) | $B_y$ | 2 | - | 4 | - |
| Inflow (outflow) velocity (km/s) | $V_x$ ($V_y$) | 50 | (100) | 50 | (100) |
| Alfvén speed (km/s) | $V_A$ | 22 | - | 35 | - |
| Ion sound speed (km/s) | $c_S$ | 18 | 40 | 18 | 32 |
| Dynamic Beta | $\beta_{dyn}$ | 10 | - | 4 | - |
| Thermal Beta | $\beta_{th}$ | 1.1 | - | 0.4 | - |
| Lundquist number | S | - | 14 | - | 18 |
| Layer half-length (mm) [i] | $L = R_C/2$ | - | 7 | - | 7 |
| Layer half-thickness (mm) | $\delta$ | - | 0.3 | - | 0.3 |
| Ion skin depth (mm) | $d_i = c/\omega_{pi}$ | 0.89 | 0.37 | 0.71 | 0.33 |
| Ion-ion mean free path (mm) | $\lambda_{ii}$ | $10^{-3}$ | $10^{-2}$ | $10^{-3}$ | $10^{-2}$ |
| Radiative cooling time (ns) | $\tau_{rad}$ | 23 | 5 | 15 | 4 |
| Ion-electron energy exchange time (ns) | $\tau_{ei}^E$ | 50 | 40 | 30 | 20 |

[i] The length $R_C$ is the radius of curvature of the magnetic field lines at the boundary of the reconnection layer.

analysis of the measured plasma parameters (Table I) indicates that the required pressure balance for this is indeed accounted for at both the early and late times. During the early stage of the experiments, when the ion temperature of the layer is large, the ram pressure of the flow exactly matches the thermal pressure of the layer. Later in time, the pressure balance is achieved between the inflow ram pressure and the magnetic pressure of the field enhancements at the layer boundary.

The Lundquist number calculated for the system is $S = LV_A/D_M \sim$10-20, where $V_A$ is the upstream Alfvén velocity, $D_M$ the magnetic diffusivity, and the length scale L of the system is defined as half the radius of curvature ($R_C$) of the azimuthal magnetic field lines at the mid-plane between the two wire arrays (Fig. 1(a)). Combining this estimate with the ratio of the length scale to the ion skin depth ($L/d_i \sim$20) leads to the expectation that the system should lie in the "single x-line collisional reconnection" domain[24], where the reconnection layer is not expected to be unstable to tearing mode (plasmoid) instabilities[38,39]. It is important to note however that this comparison to the known phase-space does not consider the super-Alfvénic nature of the inflows, which could have consequences upon this behavior. Nevertheless, the prediction appears consistent with the observations of the reconnection layer structure, which show that despite the presence of density modulations in the inflowing plasma (Fig. 3(a)-(d)), the layer is highly symmetric about its center, and displays smoothly varying density and velocity profiles along its length (Figs. 2, 3 and 7(b)). In contrast, similar experiments carried out using this pulsed power platform, but employing a carbon plasma, with a dynamic Beta parameter $\beta_{dyn} \approx 1$ and sub-Alfvénic inflow velocity, display a much more unstable reconnection process[21–23]. The carbon reconnection layer has a Lundquist number of S≈100 due to a higher electron temperature, caused by the absence of radiative cooling. This likely places the carbon experiments in the "semi-collisional" reconnection regime[39], and plasmoids are observed forming and propagating throughout the layer structure. These observed differences demonstrate the versatility of the pulsed power setup, as it can access different regimes of reconnection physics with the available control over the plasma material. Further details surrounding the tuneability of the setup and how Lundquist number parameter space can be explored are discussed in Ref. 23.

### B. Magnetic flux annihilation and plasma outflows from the reconnection layer

Faraday rotation measurements show evidence of magnetic field accumulating in pile-up regions at the reconnection layer boundaries. To determine whether this could significantly reduce the rate of magnetic flux annihilation inside the layer, the rate at which magnetic field builds up at the boundaries is compared to the rate of inflow of magnetic flux. The magnetic flux inflow rate is given by the upstream product $B_y V_x$, and the pile-up rate is estimated as the rate of growth of the field enhancements $dB_{pile}/dt$ (measured from the evolving magnetic field distribution, e.g. Figs. 4(d), 4(f) and 4(h)) multiplied by the enhancement thickness $\Delta x$. The ratio equates to:

$$\left(\frac{dB_{pile}}{dt}\Delta x\right)\bigg/(B_y V_x) \sim 10\%.$$

Within the resolution of the measurements, this suggests that the majority of the magnetic flux passes through the pile-up region and is processed inside the layer. However, a higher magnitude of $B_{pile}$ at the sharp peaks of the pile-up region cannot be ruled out due to the line averaged nature of the magnetic field measurements with the Faraday rotation diagnostic.

The destruction of magnetic flux in the reconnection layer leads to plasma heating and the formation of fast, symmetric outflows of plasma along the layer. This bulk plasma motion is consistent with the acceleration of material in the direction of the expected magnetic tension force of the reconnected field lines. The plasma outflow is measured to reach velocities of $V_y \gtrsim 100$ km/s $\sim 4V_A$, with respect to the upstream Alfvén speed. This is consistent with the generalized Sweet-Parker model of Refs. 40,41, as the outflow is able to acquire this super-Alfvénic velocity from the additional acceleration of the thermal pressure gradient in the direction of the open downstream boundary, i.e. vacuum.

In addition to motion in the xy-plane, results from TS measurements indicate the presence of a significant ion motion in the vertical out-of-plane, z-direction (Fig. 7(b)). This $V_z$ velocity of $\sim 50$ km/s, measured in the middle of the layer, is in the direction of the reconnection electric field ($E_z = J_z/\sigma$). It is interesting to compare this unexpectedly high velocity with the drift velocity between the electrons and ions required to support the current in the reconnection layer. A current density of $\sim 0.5$-1 MA/cm$^2$ is required to provide a Harris-like magnetic field profile with the measured half-thickness $\delta$ and upstream field strength. Combined with the measured electron density, this current density corresponds to a drift velocity of $U_d = J_z/n_e = 25$-50 km/s. This is comparable to the measured ion velocity, suggesting that the vertical ion motion could make a considerable contribution to the current inside the reconnection layer. This raises the intriguing possibility of the ions acting as the primary charge carrier responsible for the current, and therefore merits a future investigation of the out-of-plane velocity distribution of the reconnection layer.

### C. Energy partition

Measurements from TS show that there is a clear evolution over time of the ion temperature inside the reconnection layer (Fig. 8(b)). Early in time (t=215 ns) $T_i \gg T_e$ in the layer, with $T_i \approx \bar{Z}T_e \sim 300$ eV. It can be demonstrated that this high ion temperature, which is an order of magnitude greater than the temperature of the upstream flow, exceeds what can be expected by both strong-shock heating from the supersonic entry into the layer, and viscous heating due to the high velocity shear between the layer and upstream plasma.

The 50 km/s inflow velocity measured upstream of the layer boundary, where $T_i$ is small, corresponds to Al ions with a directed kinetic energy of $E_i = m_i V_x^2/2 = 350$ eV. Thermalization of this kinetic energy (assuming no energy is transferred to the electrons) gives a maximum possible ion temperature of $T_i = (2/3)E_i$, which is already smaller than the measured post-interaction $T_i$ at early time. The actual upper limit of the post-shock plasma temperature is significantly smaller, and can be estimated using a standard expression for heating in a strong shock[42],

$$k_B T_i = E_i \frac{4(\gamma - 1)}{(\gamma + 1)^2}\frac{1}{(\bar{Z} + 1)}, \quad (4)$$

where $\gamma$ is the adiabatic index of the plasma. Using $\gamma = 5/3$, and neglecting equilibration with the electrons (i.e. $\bar{Z} = 0$), gives an upper limit for the immediate post-shock ion temperature of $T_i = 120$ eV. This reduces to $T_i$=30-15 eV for $\bar{Z}$=3-7, once ion-electron equilibration is established.

The viscous heating rate can be estimated by considering the viscous damping of the highly sheared velocity profile of the outflows. Following the treatment of Ref. 43 and employing Braginskii's expression for the ion viscosity[44]:

$$\frac{3}{2}n_i k_B \frac{\partial T_i}{\partial t} = 0.96 n_i k_B T_i \tau_i \left(\frac{\partial V_y}{\partial x}\right)^2, \quad (5)$$

where $\tau_i \propto T_i^{3/2}$ is the ion collisional timescale. Solving this differential equation using the parameters in Table I and assuming a maximum velocity shear, where $V_y$ drops from 100 km/s at the center of the layer to zero at $|x| = \delta$, gives a viscous heating timescale $\gg$500 ns for even a modest heating to 100 eV from the initial 30 eV of the upstream flow.

Thus, an additional mechanism for ion heating must be present inside the layer, which should be expected to draw from the released magnetic energy. Enhanced heating of ions has been discussed extensively in the context of magnetic reconnection, e.g. in Refs. 12,43,45,46, and is often associated with the development of kinetic plasma turbulence. The high current density at the boundary of the current layer corresponds to a drift velocity exceeding the ion sound speed ($U_d/c_S \sim 5$). This could lead to the development of e.g. ion-acoustic or lower hybrid drift instabilities[13], which may be detectable with Thomson scattering measurements[47,48], but additional experiments would be needed to investigate this further.

At late times in the experiments (t$\gtrsim$240 ns), the thermal properties of the layer reach an approximately steady state, with the ion temperature in the layer roughly equal to that of the electrons, at values comparable to those predicted above from Eq. (4) ($T_i \approx T_e \sim 30$ eV). Calculating the energy exchange time between these populations ($\tau_{ei}^E \propto 1/n_i$) shows that as the ion density of the layer increases from $n_i = 2 \times 10^{17}$ cm$^{-3}$ at t=215 ns, to $n_i = 4 \times 10^{17}$ cm$^{-3}$ at t=250 ns, the exchange time drops over the range $\tau_{ei}^E = 40 \to 20$ ns. In comparison, the time taken for inflowing plasma to exit the central region of the layer at the measured outflow velocity is of the order $\sim$50-100 ns. This indicates that as the experiment progresses the system converges towards a situation where the ion and electrons have time to equilibrate before the plasma leaves the layer.

The electron temperature is approximately constant throughout the experiments, and thus the internal energy $U_{int}$ of the electron population should be conserved by a balance of the in- and outgoing energy fluxes. The incoming energy can be evaluated as the sum of the contributions from the ion-electron exchange ($(3/2)n_e k_B (T_i - T_e)/\tau_{ei}^E$) and the resistive heating of the electrons (estimated classically from the Spitzer resistivity as $\eta_{Sp} J_z^2$). These must counter the radiative cooling losses of the plasma, which can be represented as a cooling function $\Lambda(n_i, T_e)$. Calculations of $\Lambda$ for aluminum, following the approach described in Refs. 49,50, show that the radiative power loss at the relevant conditions of the layer is significant, and this is reflected by the relatively short cooling timescales ($\tau_{rad} = U_{int}/n_e n_i \Lambda$) quoted for the layer in Table I. In comparison to the radiative power loss, the heating power provided by the ion-electron exchange and the resistive heating inside the reconnection current sheet (assuming a Harris-like profile) accounts for only $\sim$50% of the required energy input to the electrons to keep an approximately constant temperature at both early and late times (the ratio however shifts from $\sim$40% of the required energy input being provided by ion-electron exchange, and $\sim$10% from resistive heating, at early time, to the reverse situation at late time). The most plausible explanation for this apparent shortfall in heating power is that an additional resistive heating is provided by the current sheets associated with the magnetic flux pile-up at the boundaries of the layer. The large spatial gradients of the magnetic field seen here correspond to current densities growing from $\sim$3-6 MA/cm$^2$ over the range t=215-250 ns. Thus, this could potentially provide a short but intense heating power to the electrons as the plasma passes across the layer boundary, however higher resolution TS measurements would be required to verify this hypothesis.

## V. CONCLUSIONS

This paper describes the structure and evolution of a long-lasting reconnection layer formed by colliding magnetized, aluminum plasma flows in the strongly-driven regime (high ratio of ram to magnetic pressure, super-Alfvénic inflow velocity). The reconnection layer is dynamically stable, highly symmetric and quasi-two-dimensional, allowing ease of access for diagnosis of the spatially and temporally resolved plasma parameters of the system. The boundary conditions set by the driven inflow result in a reconnection layer which shows evidence of a strong pile-up of the magnetic flux brought by the inflows at the boundaries of the layer. Early in time the reconnection layer shows an unexpectedly large ion temperature, $T_i \approx \bar{Z} T_e \gg T_e$, which cannot be explained by considerations of either the thermalization of the kinetic energy of the inflowing material, or by classical viscous heating. Later in time the ions inside the layer cool, via an increased rate of energy exchange to the electrons. Meanwhile, the electron population maintains an approximately constant temperature throughout the lifetime of the reconnection layer, via an evolving balance between the heating contributions from the ions and resistive heating, and the strong losses due to radiative cooling of the aluminum plasma.

A number of significant differences are found between both the structures and thermal properties of the reconnection layer observed in these experiments, and in experiments using a geometrically identical setup but with sub-Alfvénic, carbon plasma inflows[21–23]. In both experiments the reconnection layer forms with a Harris-like magnetic field profile. However, only in aluminum, with its much higher dynamic Beta parameter, does this profile later evolve to show the strong field enhancements associated with magnetic field pile-up. The strong radiative cooling of the aluminum plasma also plays a large role in the differences of these systems. The radiative cooling results in a much lower electron temperature inside the aluminum reconnection layer, and consequently the system has a much smaller Lundquist number (S$\sim$10 in aluminum, versus $\sim$100 in carbon). This appears to prevent tearing-mode instabilities in the aluminum reconnection layer, allowing single-x-line reconnection layer to operate, without the formation of plasmoids, such as those observed in the carbon reconnection experiments[21–23]. These differences highlight the suitability of this pulsed power setup for studying reconnection processes under a range of conditions and parameter space.

## ACKNOWLEDGEMENTS

This work was supported in part by the Engineering and Physical Sciences Research Council (EPSRC) Grant No. EP/N013379/1, and by the U.S. Department of Energy (DOE) Awards No. DE-F03-02NA00057, DE-SC-0001063 and DE-NA-0003764. A.C. and N.F.L. were supported by LABEX Plas@Par with French state funds managed by the ANR within the Investissements d'Avenir programme under reference ANR-11-IDEX-0004-02. N.F.L. was supported by the NSF-DOE partnership in Basic Plasma Science and Engineering, Award No. DE-SC-0016215.